\documentclass[twocolumn,amsmath,amssymb,floatfix]{revtex4}
\usepackage{graphicx}

\begin{document}

% \draft

\title{Three-terminal devices to examine single molecule conductance switching}

\author{Z.K. Keane$^{1}$, J.W. Ciszek$^{2}$, J.M. Tour$^{2,3}$, D. Natelson$^{1,4}$}

\affiliation{$^{1}$ Department of Physics and Astronomy, $^{2}$ Department of Chemistry, $^{3}$ Department of Computer Science and the Smalley Institute for Nanoscale Science and Technology, $^{4}$ Department of Electrical and Computer Engineering, Rice University, 6100 Main St., Houston, TX 77005}

\date{\today}

\begin{abstract}

We report electronic transport measurements of
single-molecule transistor devices incorporating bipyridyl-dinitro
oligophenylene-ethynylene dithiol (BPDN-DT), a molecule known to
exhibit conductance switching in other measurement configurations.
We observe hysteretic conductance switching in 8\% of devices with
measurable currents, and find that dependence of the switching
properties on gate voltage is rare when compared to other
single-molecule transistor devices.  This suggests that polaron
formation is unlikely to be responsible for switching in these
devices.  We discuss this and alternative switching mechanisms.
\end{abstract}

% \pacs{73.22.-f,73.23.Hk,85.65.+h}
\maketitle

%\newpage

One area of strong interest within molecular electronics has been
molecular switching, the experimental observation that conduction
through some molecular systems has been observed to switch
discretely between states of comparatively high and low conductance.
Discrete switching has been observed in a number of measurement
configurations and molecules\cite{MooreetAl06JACS}. Scanning
tunneling microscopy (STM) has been particularly useful in studying
this phenomenon by observing molecules of interest assembled at
grain boundaries and defects in alkane self-assembled monolayers
(SAMs).  Of particular interest is voltage-driven switching, when
sweeping a dc bias voltage beyond a threshold $V_{\rm on}$ triggers
the transition to the higher conducting state.  This enhanced
conduction persists until the bias
is reduced to %values below $V_{\rm on}$,
$V_{\rm off} < V_{\rm on}$
leading to hysteretic current-voltage characteristics.  A recent
paper\cite{BlumetAl05NM} examined this phenomenon in bipyridyl-dinitro
oligophenylene-ethynylene dithiol (BPDN-DT) using STM, a crossed wire
method\cite{KushmericketAl02PRL}, and a nanoparticle-based
technique\cite{LongetAl05APL}, finding similar switching
characteristics in all three measurement approaches.  This suggests
that the switching in this case is intrinsic to the molecule/metal
system, rather than an artifact of a particular measurement
technique.

One suggested switching mechanism that has both motivated molecular
design and been the focus of intense theoretical analysis is reduction
of functional groups on the molecule\cite{SeminarioetAl02JCP}.
Polaron formation is one way of potentially stabilizing this
reduction\cite{GalperinetAl05NL}.  Strong coupling between an
electronic level and a vibrational mode localized to the molecule can
lead to a renormalization of that level to a lower energy when
occupied.  In the context of switching, the physical idea is that
large source-drain bias voltages can alter the average electronic
population of the molecule from its zero bias value, and that excess
charge can be stabilized via this polaronic process, remaining on the
molecule as bias is then reduced.  The stability of polarons depends
in detail on the local electronic environment.  There have been numerous
theoretical examinations of such a
model\cite{AlexandrovetAl03PRB,MitraetAl05PRL,GalperinetAl05NL,MozyrskyetAl06PRB},
with several reporting that bistable and hysteretic conduction can
arise with certain ranges of parameters.
Other proposed switching mechanisms include macroscopic changes in
molecular conformation\cite{CollieretAl01JACS}, rotation of functional
groups\cite{diVentraetAl01PRL} or conjugated
rings\cite{CorniletAl02JACS}, bond fluctuation at the attachment point
between the molecule and the
electrode\cite{RamachandranetAl03Science}, and changes in
hybridization between the molecule and the
electrode\cite{KornilovitchetAl01PRB,DonhauseretAl01Science,LewisetAl05JACS,MooreetAl06JACS}.

Single-molecule transistors (SMTs) can be used to examine candidate
switching mechanisms.  SMTs are three-terminal devices with conduction
occuring between source and drain via a single small molecule,
modulated by capacitive coupling to a proximal gate
electrode\cite{Natelsonrev}.  SMTs have been used to study other
vibrational
phenomena\cite{ParketAl00Nature,YuetAl04NL,YuetAl04PRL,PasupathyetAl05NL,ChaeetAl06NL,vanderZant06FD},
and capacitive coupling to the gate has been used to manipulate the
average electronic occupation of the molecules.  As the gate voltage, $V_{\rm G}$,
is increased, the electrostatic interaction with the gate tends
to favor an increased electronic population of the molecule.

We report measurements of SMTs incorporating BPDN-DT.  We find
hysteretic conductance at high source-drain voltages, $V_{\rm SD}$,
qualitatively consistent with the observations of Blum {\it et
al.}\cite{BlumetAl05NM}.  Control measurements on SMT structures
incorporating only alkane chains show no such switching.  Gate
dependence in BPDN-DT devices is rare, compared with thousands of SMTs
previously made from C$_{60}$ and transition metal complexes.
We discuss the implications of this for the switching mechanism
at work in this molecule.

The fabrication of SMTs has been discussed extensively
elsewhere\cite{YuetAl04NL,YuetAl04Nanotechnology}.  Using e-beam
lithography, we define constricted wires on [100] \emph{p+} Si wafers
coated with 200~nm of thermally grown SiO$_{2}$.  Each wafer is
patterned with at least 45 devices.  Source and drain electrodes are
1~nm Ti/20~nm Au deposited by e-beam evaporation.  After lift-off and
O$_{2}$ plasma cleaning, devices are immersed in a 0.2~mM solution of
acetate protected BPDN-DT\cite{FlattetAl03Tetrahedron} in a
nitrogen-purged 1:1 mixture of THF and ethanol.  Thiol-based
self-assembly is carried out in the standard alkaline deprotection
chemistry\cite{CaietAl02CM} for 48~h.  The stability of the NO$_{2}$
groups during self-assembly is demonstrated in Supporting Information.
Finally, the gold wires are electromigrated\cite{ParketAl99APL} to
failure at 4.2~K, and electrical transport measurements are made as a
function of source-drain bias, $V_{\rm SD}$, and gate voltage, $V_{\rm
G}$ at 10~K in a variable temperature vacuum probe station. The
cryogenic environment minimizes the risk of adsorbed contaminants, and
increases device geometric stability by inhibiting diffusion of metal
atoms and molecules.  DC measurements of device current-voltage
characteristics ($I_{\rm D}-V_{\rm SD}$) are performed with the source
electrode grounded, at various $V_{\rm G}$.  Initial
post-electromigration characterization is via a source-drain bias
sweep up to 100~mV.

The nanoscale variation between devices produced by the
electromigration method necessitates a statistical approach to
device characterization.  Of 464 devices fabricated in this manner
incorporating BPDN-DT, 169 exhibited no measurable tunneling current
after electromigration.  Previous experience and SEM imagery suggest
that the most likely explanation is a resulting source-drain gap
several nm or larger.  All remaining devices found to have
detectable conduction were measured at source-drain biases as high
as 2~V.

These voltages are significantly higher than those used in previous
experiments\cite{YuetAl04NL,YuetAl04PRL,YuetAl05PRL}, and are
necessary to search for switching phenomena like those reported for
this molecule in other geometries\cite{BlumetAl05NM}.  For gaps in the
range of 2~nm, the resulting source-drain electric fields at high bias
can readily exceed the breakdown threshold of many materials.  No sign
of arcing or destructive irreversibility is observed in these devices
after voltage application.  However, a common failure mode of SMTs
produced by electromigration is device instability: conductance
characteristics that change irreversibly in the course of systematic
measurements, presumably due to alteration of the device geometry
(specifically the rearrangement of electrode atoms, or the breaking of
the molecule/electrode bond).  This instability is likely to be driven
by both the electric field and the current density at the gap.  These
irreversible changes, which almost always result in lower overall
conductance, make long studies at high biases very difficult.

\begin{figure}[h!]
\begin{center}
\includegraphics[clip, width=8cm]{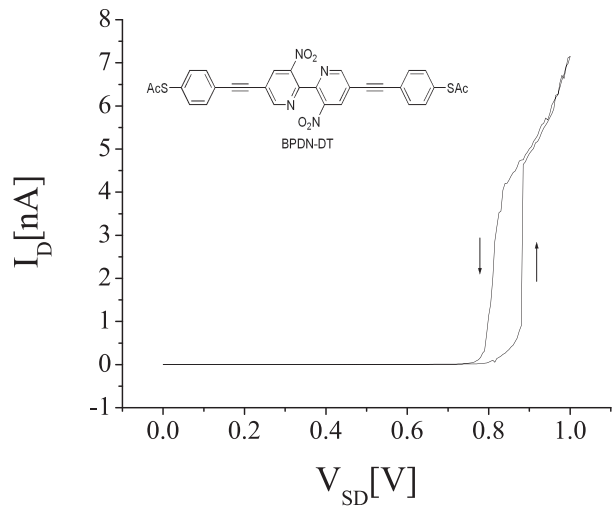}
\end{center}
\vspace{-5mm} \caption{\small{I-V curve of a typical BPDN-DT device
exhibiting conductance switching and hysteresis.  Here, $V_{\rm on}
= 810~mV$ and $V_{\rm off} = 885~mV.$  Inset: Structure of the
BPDN-DT molecule.}} \label{fig1}
%\vspace{-5mm}
\end{figure}

Out of the devices with measurable conduction, 24 exhibited hysteretic
behavior qualitatively like that shown in Fig. 1.  Additional
representative curves are shown in Supporting Information, to give a
sense of the device-to-device variation.  The source-drain bias is
swept from zero to a pre-defined endpoint (typically 1~V) in 40~s,
then back to zero.  Conductance is low ($< 10^{-8}$~S) until a
characteristic threshold $V_{\rm SD} \equiv V_{\rm on}$ is reached.  A
discontinuous transition to a higher conductance state occurs; some
devices exhibit regions of astability, in which several rapid
transitions between the high and low conductance states occur, for
$V_{\rm SD}$ just exceeding $V_{\rm on}$.  At higher biases, the
device remains in the high-conductance state.  As the bias is swept
back down, the device remains in the high-conductance state until a
second characteristic bias, $V_{\rm off} < V_{\rm on}$ is reached, at
which point there is a transition back to the low-conductance
state. In two additional devices, hysteresis was also observed, but
with the opposite sense; that is, initial sweeps began in a more
conducting state and at high biases a transition was observed to a
less conducting state.

Control experiments were performed using both bare junctions and
electromigrated junctions incorporating dodecanethiol ($C_{\rm
12}SH$). Out of a total of 145 control device subjected to the same
biasing scheme, no devices exhibited this hysteretic conduction.
This is consistent with the BPDN-DT molecules being responsible for
the hysteresis.

\begin{figure}[h!]
\begin{center}
\includegraphics[clip, width=8cm]{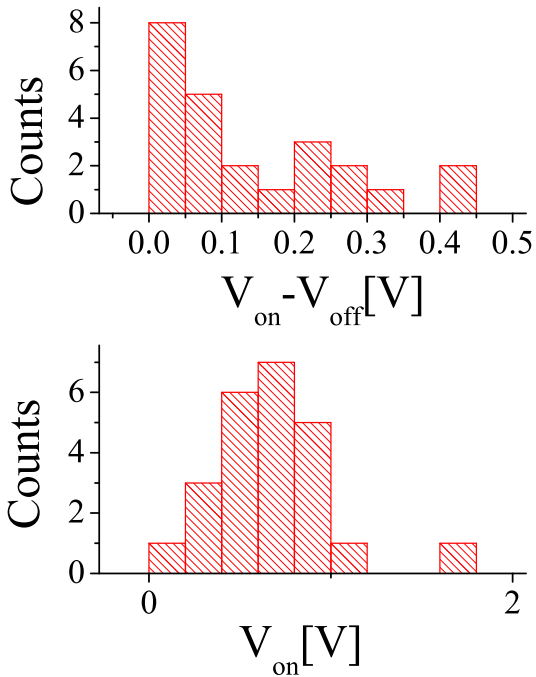}
\end{center}
\vspace{-5mm}
\caption{\small{Lower: Histogram showing the wide variability in
$V_{on}.$
Upper: Histogram showing the variability in $V_{on} -
V_{off}.$}
} \label{fig2}
%\vspace{-5mm}
\end{figure}

Both the switching voltages and the width (in $V_{\rm SD}$) of the
hysteretic region vary considerably from device to device.  As shown
in Fig. 2, $V_{\rm on}$ ranges from 390~mV to as high as 1.7~V, but
in most of these devices is around 700~mV.  This correlates well
with previously reported results on voltage-controlled conductance
switching in BPDN-DT\cite{BlumetAl05NM}.  The astability near the
transition points between the two conductance branches further
complicates attempts to quantify the width of the hysteresis. A
hysteretic device under typical bias sweep conditions survives on
the order of 10 bias sweeps (all showing hysteresis) before
irreversibly changing to a lower conductance, non-hysteretic
configuration, though occasional devices exhibited much better
stability, surviving hundreds of cycles.

In addition to the source and drain, the BPDN-DT molecules in our
devices are capacitively coupled to a gate electrode.  The effective
strength of this coupling depends strongly on the geometry of the
junction, and the strength of the coupling between the molecule and
the electrodes.  In previous SMT experiments we have examined
thousands of devices, and found that approximately 10-15\% of the
initially patterned electrodes show significant gate dependence.
This is true both for physisorbed molecules
(C$_{60})$\cite{YuetAl04NL,YuetAl04Nanotechnology} and transition
metal complexes attached via gold-thiol
bonds\cite{YuetAl04PRL,YuetAl05PRL}.  In the BPDN-DT devices
examined, only two out of the original 464 devices show any
systematic gate dependence; Fig. 3 shows a set of $I_{\rm D}-V_{\rm
SD}$ curves from one of these devices.

\begin{figure}[h!]
\begin{center}
\includegraphics[clip, width=8cm]{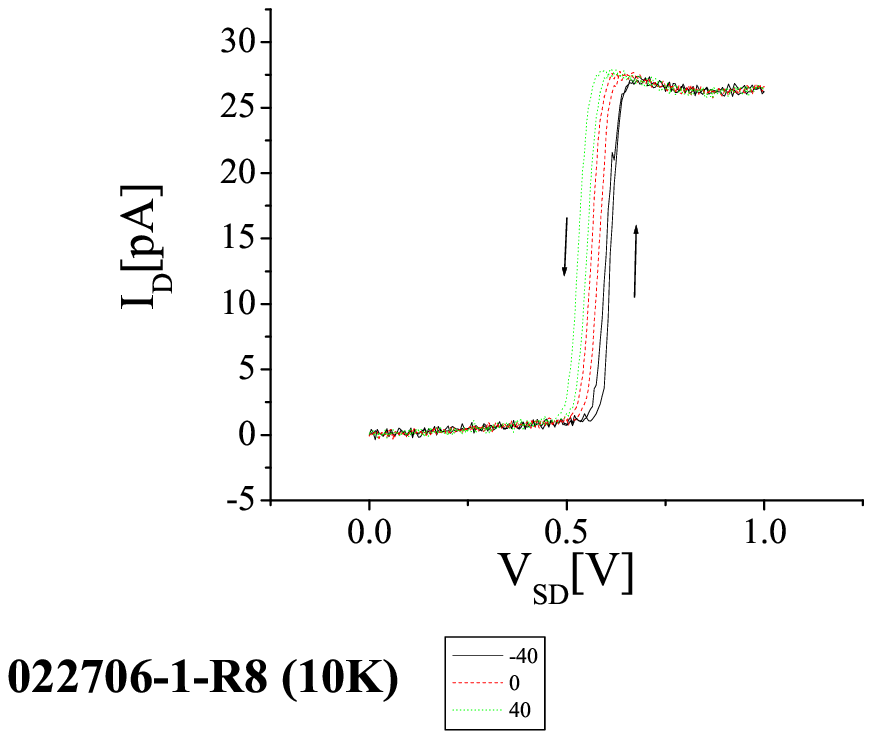}
\end{center}
\vspace{-5mm} \caption{\small{(Color online) Gate dependence of one
device. Moving from left to right, the curves represent $V_{G} =
+40$~V (green, dashed), 0 (red, dotted), and -40~V (black, solid),
respectively.} Arrows indicate the sense of the hysteresis for each
curve.} \label{fig3}
%\vspace{-5mm}
\end{figure}

Detailed studies of the gate dependence ({\it e.g.} the conductance
maps as a function of $V_{\rm SD}$ and $V_{\rm G}$ familiar from
Coulomb blockade devices) were impeded by the short device lifetime
described above.  However, it was possible on the other gate-dependent
device to sit at a bias near $V_{\rm on}$ and sweep $V_{\rm G}$, to
search for any systematic variation of $V_{\rm on}$ with gate
potential.  The data are shown in Fig. 4.  The device shows sporadic,
hysteretic switching between conductance states as $V_{\rm G}$ is
varied, with a weak, nonmonotonic trend toward lower conduction at
more positive $V_{\rm G}$.  Note that this is the {\it opposite} trend
as that seen in the device of Fig. 3.

\begin{figure}[h!]
\begin{center}
\includegraphics[clip, width=8cm]{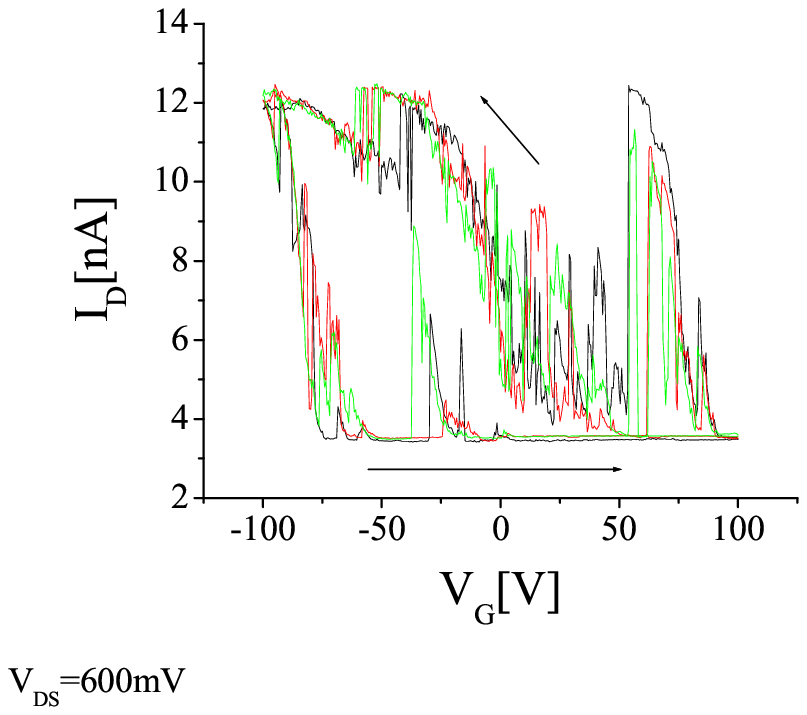}
\end{center}
\vspace{-5mm} \caption{\small{(Color online) Current through a BPDN
device as a function of $V_{\rm G}$ with $V_{\rm SD} = 600~{\rm mV}
\simeq V_{\rm on}$.  This device was measured starting at $V_{\rm G}
= +100$~V, sweeping down to -100 V and back several times.} }
\label{fig4}
%\vspace{-5mm}
\end{figure}

These data constrain the switching mechanism discussion
significantly. If the source of the hysteresis is an electronic effect
such as polaron formation, one would expect a reproducible, systematic
trend of $V_{\rm on}$ with $V_{\rm G}$.  An increasingly positive
$V_{\rm G}$ acts to stabilize additional electronic population on the
molecule, presumably lowering the source-drain bias required to
populate the next electronic level.  The efficiency of gate coupling
in previous SMT devices with this oxide
thickness\cite{YuetAl04NL,YuetAl04PRL,YuetAl05PRL} has generally been
on the order of a few percent, implying that a 100~V swing of $V_{\rm
G}$ should systematically shift molecular levels (and correspondingly
$V_{\rm on}$ and $V_{\rm off}$) by a few hundred meV.  Whether this is
sufficient to bring a molecular level into alignment with the
source/drain chemical potential depends on the details of the
molecule's electronic structure (gap between highest occupied and
lowest unoccupied molecular orbitals) and the molecule-metal bonding.
Further, while transport via highest occupied {\it vs.} lowest
unoccupied orbitals could affect the sign of the gate trend, the
differences between Figs. 3 and 4 would imply that qualitatively
identical transport in different devices takes place through different
orbitals, which seems unlikely.

Within the polaronic picture of hysteresis, which implies that the
energetic difference between the source/drain chemical potential and a
molecular orbital is $\sim eV_{\rm on}$, there are limited possible
explanations for the observed weak gate coupling.  If this particular
molecule self-assembles or moves during electromigration in a manner
very different than in other experiments, it is conceivable that there
could be a systematic trend toward device geometries with weak gate
couplings.  This seems unlikely, given the relatively ordinary
self-assembly properties of BPDN-DT reported by Blum {\it et
al.}\cite{BlumetAl05NM}.  Very strong electronic coupling of the
BPDN-DT molecules to at least one of the electrodes could also explain
a relatively weak effect of the gate.  If the molecular levels are
strongly pinned relative to the Fermi level of a lead, then
effectively the molecule will be screened well from the gate
potential.  Such a strong lead coupling would strongly constrain
dynamic polaron formation, requiring weak gate coupling, yet a large
enough reorganization energy that small bias-driven charge transfer
could shift the level into or out of resonance, all relatively
independent of the details of the Au site to which the molecule is
bound.  Detailed quantum chemistry calculations should at least be
able to assess the likelihood of such strong coupling between the
molecule and the metal, as well as the level alignments mentioned
above.

A more plausible explanation, in our view, is that the switching
mechanism in these devices is based on bias-driven changes in the
molecule/electrode hybridization or bond angle, rather than a direct
electronic process like polaron formation.  The lack of gate
dependence would then naturally follow from a large energetic
difference between molecular levels and the source/drain chemical
potential.  Given the heterogeneous environment of the electromigrated
junctions, it would not be surprising for the molecule to bind
asymmetrically to the electrodes.  Indeed, during initial SAM
formation, only one end of the molecule is expected to be bound to the
Au surface.  Under this circumstance, the bound molecule can acquire a
net electric dipole moment, which may be enhanced by the strong
electron-withdrawing character of the nitro groups in BPDN-DT.  STM
experiments have already demonstrated\cite{LewisetAl05JACS} that
electric dipoles can lead to a bias influence on conductance switching
in other molecules, which has been interpreted as due to changes in
molecule/electrode
hybridization\cite{LewisetAl05JACS,MooreetAl06JACS}.

We have performed measurements of electromigrated junctions in a SMT
configuration incorporating BPDN-DT, a molecule known from other
experiments to exhibit hysteretic conductance switching as a function
of $V_{\rm SD}$.  We observe similar hysteretic characteristics that
are absent in control devices not incorporating the molecule of
interest.  The dependence of the conductance on $V_{\rm G}$ is much
weaker than observed with other molecules in SMT experiments. While
not the only possibility, a
likely explanation is that the switching mechanism in these devices
is not charge-based, but rather depends on the geometry of the
molecule/metal interface.  Performing SMT experiments using molecules
known to have strong polaron formation tendencies remains an
interesting direction to pursue.  Similarly, because polaron stability
depends strongly on the electronic environment, examination of
switching processes in electrochemical cells ({\it i.e.} immersed in
electrolyte) are also of much interest.

DN acknowledges support from the Research Corporation, the Robert A.
Welch Foundation, the David and Lucille Packard Foundation, an
Alfred P. Sloan Foundation Fellowship, and NSF award DMR-0347253.
JMT acknowledges support from DARPA via the AFOSR.

%\clearpage

%\clearpage

%\clearpage

%\clearpage

%\clearpage

\end{document}